\documentclass[11pt]{article}
\usepackage[preprint]{acl}
\usepackage{marvosym}
\usepackage{times}
\usepackage{latexsym}
\usepackage[T1]{fontenc}
\usepackage[utf8]{inputenc}
\usepackage{microtype}
\usepackage{graphicx}
\usepackage{booktabs}
\usepackage{amsmath,amssymb}
\usepackage{multirow}
\usepackage{url}
\usepackage[most]{tcolorbox}
\usepackage{tabularx}
\usepackage{longtable}
\usepackage{xcolor}
\usepackage{colortbl}
\definecolor{TableHeader}{HTML}{F3F5F6}
\definecolor{TableBand}{HTML}{E2F0F7}
\newtcolorbox{promptbox}[1]{
  colback=gray!6,
  colframe=gray!45,
  boxrule=0.4pt,
  arc=0mm,
  breakable,
  title={#1},
  fonttitle=\small\bfseries,
  coltitle=black,
  colbacktitle=gray!12,
  fontupper=\footnotesize\ttfamily,
  left=1.5mm,
  right=1.5mm,
  top=1.2mm,
  bottom=1.2mm
}

\title{Beyond the Vote: When Do LLMs Add Value to Retrieval-Based \texorpdfstring{\\}{ }Tabular Classification?}

\author{
    Ziquan Wang\textsuperscript{1} \quad
    Yuxuan Nie\textsuperscript{2} \quad
    Jian Ma\textsuperscript{1} \quad
    Yingjie Li\textsuperscript{1}\thanks{Corresponding author.}
    \\[0.7em]
    \textsuperscript{1}Hangzhou International Innovation Institute, Beihang University, Hangzhou, China
    \\
    \textsuperscript{2}Peking University, Beijing, China
    \\[0.3em]
    {\itshape
    \{ziquanwang,yingjieli\}@buaa.edu.cn}
}

\hypersetup{
  pdfauthor={Ziquan Wang, Yuxuan Nie, Jian Ma, Yingjie Li},
  pdftitle={Beyond the Vote: When Do LLMs Add Value to Retrieval-Based Tabular Classification?},
  pdfsubject={Preprint}
}

\begin{document}
\setcounter{footnote}{1}
\maketitle
\maketitle

\suppressfloats[t]
\begin{abstract}
Retrieval-based in-context learning (ICL) enables large language models (LLMs) to predict tabular labels from similar examples, but it remains unclear whether LLM inference improves on direct voting over those same examples. We address this question by comparing four general-purpose LLMs with distance-weighted nearest-neighbor voting (Distance-KNN) using identical retrieved neighbors across 19 tabular classification datasets. With the number of demonstrations fixed, enlarging the retrieval candidate pool improves accuracy for all predictors, yet Distance-KNN achieves the highest mean accuracy at every pool size. More detailed analyses reveal that LLM gains are concentrated where retrieved neighbors exhibit low label consensus and vary across models. These gains further depend on the average feature-space distance to the test query: under low label consensus, some LLMs outperform Distance-KNN in close and intermediate neighborhoods, but all fall behind on average in the most distant group. Complementary experiments on the number of demonstrations, prompt formulation, retrieval representation, and label-balanced demonstration selection largely preserve the overall findings. These results characterize when LLM inference adds predictive value beyond distance-weighted voting in retrieval-based tabular ICL.
\end{abstract}

\begin{figure}[t]
\centering
\includegraphics[width=0.9\columnwidth]{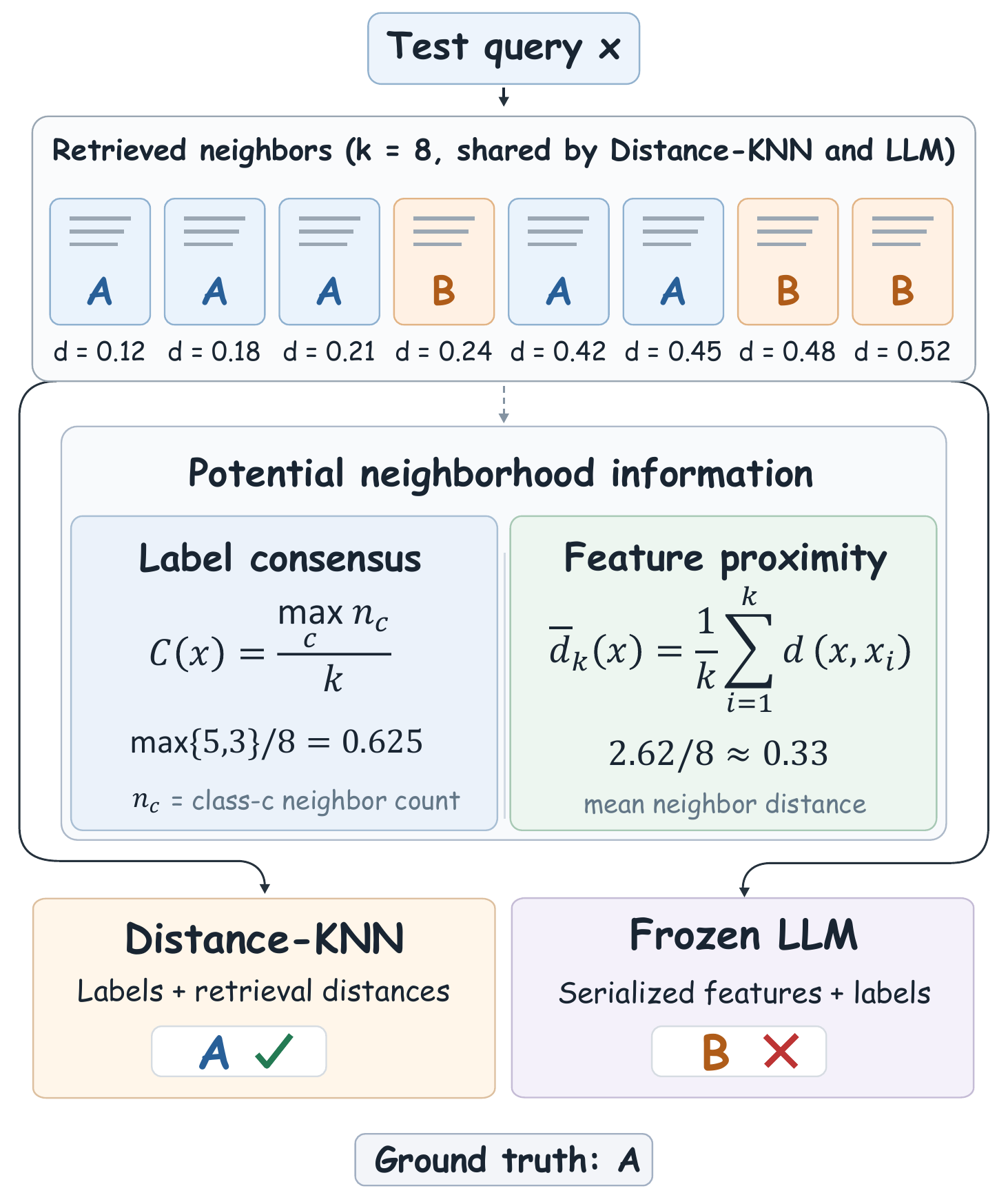}
\vspace{-1mm}
\caption{Illustration of shared-neighbor prediction in tabular classification. Distance-KNN and the LLM use the same eight retrieved neighbors. Label consensus measures the majority-label fraction, while feature proximity is measured by the mean query-to-neighbor distance. In this example, Distance-KNN correctly predicts class A, whereas the LLM predicts class B.}
\label{fig:teaser}
\end{figure}

\section{Introduction}
Tabular classification is a widely studied machine learning task with applications in healthcare \citep{yang2024ehr}, commerce \citep{sun2024sales}, and energy \citep{alawi2024energy}. The goal is to predict class labels from instances described by numerical and categorical features. Traditional methods include nearest-neighbor classifiers, which predict labels from similar training instances \citep{dudani1976}, and tree-based ensembles, which have demonstrated strong performance on heterogeneous tabular data \citep{chen2016,grinsztajn2022}. 

Recent research has explored alternative approaches that leverage pretrained models for tabular prediction. Specialized tabular foundation models, including TabPFN, learn transferrable predictive patterns through pretraining \citep{hollmann2023,hollmann2025,ma2025tabdpt}. Another line of work adapts language models to tabular data through supervised or instruction tuning \citep{hegselmann2023,li2024tablegpt,wen2024supervised,zhang-etal-2024-tablellama}. More recently, researchers have explored applying frozen general-purpose large language models (LLMs) through few-shot in-context learning (ICL) \citep{brown2020language,wu-hou-2025-efficient,yuan-etal-2025-summarize}. In this setting, labeled tabular instances are serialized as demonstrations, often selected according to their similarity to a test query, enabling prediction without task-specific parameter updates.

Despite these advances, it remains unclear how much predictive value LLM inference adds beyond direct voting on the same retrieved examples. In retrieval-based tabular ICL, performance differences may arise from either the selection of demonstrations or the predictions made from them, making these contributions difficult to distinguish when different retrieved examples are used. Moreover, expanding the candidate pool changes the available retrieval evidence, but whether LLM inference benefits more from this change than direct voting remains unclear. Their relative performance may also depend on two characteristics of the retrieved neighborhood: label consensus, the fraction of neighbors sharing the most frequent label, and feature proximity, measured by the mean feature-space distance from the test query to those neighbors, as illustrated in Figure~\ref{fig:teaser}. A controlled comparison on identical retrieved neighbors is therefore needed to assess when LLMs can provide additional predictive value.

In this work, we systematically compare four general-purpose LLMs with distance-weighted nearest-neighbor voting (Distance-KNN) across 19 tabular classification datasets using a shared-neighbor evaluation protocol. At each candidate-pool size, both predictors use the same eight retrieved examples: LLMs predict from their serialized features and labels, while Distance-KNN predicts from their labels and retrieval distances. By expanding the candidate pool while keeping the number of retrieved examples fixed, we examine their performance and how LLM gains vary with label consensus and feature proximity. Our results show that Distance-KNN achieves the highest mean accuracy across all evaluated candidate pools, while LLMs exhibit selective gains under low label consensus. Feature proximity further distinguishes these gains, with improvements occurring in selected close and intermediate neighborhoods but disappearing in the most distant ones.

In summary, our contributions are as follows: 
\textbf{1)} We present a systematic shared-neighbor comparison of four frozen LLMs and Distance-KNN across 19 datasets, enabling a direct assessment of the contribution of LLM inference over distance-weighted voting.
\textbf{2)} We find that Distance-KNN achieves the highest mean accuracy across all evaluated candidate pools, while LLMs exhibit selective gains in low-consensus neighborhoods that vary with feature proximity.
\textbf{3)} We further examine how the number of demonstrations, prompt formulation, retrieval representation, and label-balanced demonstration affect the comparison, finding that the advantage of Distance-KNN largely persists across these experiments.

\section{Related Work}
\label{sec:related}

\paragraph{Tabular Prediction and Tabular Foundation Models}
Tabular prediction has traditionally relied on distance-based methods such as k-nearest neighbors (KNN) and tree-based ensembles, including Random Forest, XGBoost, and CatBoost \citep{cover1967,breiman2001,chen2016,prokhorenkova2018}. To circumvent task-specific training while achieving strong generalization across unseen datasets, recent work has introduced tabular foundation models. For instance, TabPFN is pretrained on synthetic tabular tasks to make predictions directly from labeled examples \citep{hollmann2023,hollmann2025}. Extending this paradigm to real-world data, TabDPT introduces a dedicated tabular architecture for in-context prediction, rather than adapting general-purpose LLMs \citep{ma2025tabdpt}.

\paragraph{Adapting Language Models for Tabular Prediction}
A common paradigm relies on additional training to adapt language models for tabular prediction. Existing methods fine-tune LLMs on serialized rows for few-shot classification \citep{hegselmann2023}, train on synthetic table tasks \citep{li2024tablegpt}, or leverage diverse tabular datasets for cross-dataset transfer \citep{wen2024supervised,gardner2024}, while others explore retrieval-guided tuning \citep{wen2025scalable} and structure-aware parameter adaptation \citep{he-etal-2025-tablelora}. In contrast, our study investigates tabular prediction using frozen, general-purpose LLMs without additional task-specific training, leveraging Distance-KNN to provide a controlled comparison on identical retrieved demonstrations.

\begin{figure*}[t]
\centering
\includegraphics[width=0.9\textwidth]{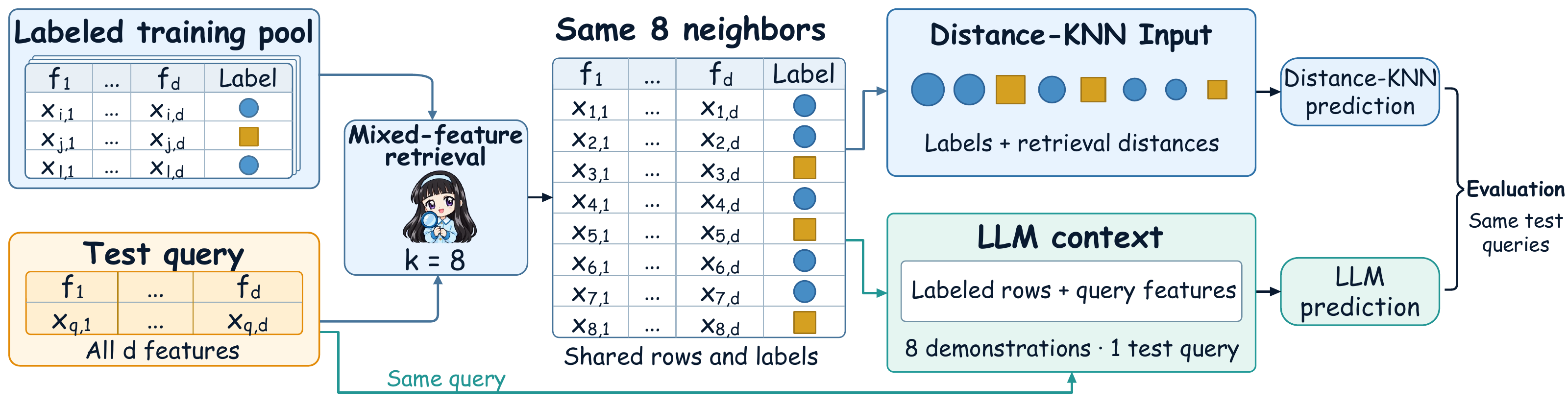}
\caption{Shared-neighbor comparison of Distance-KNN and LLMs for tabular classification. For each test query, a common retriever selects $k=8$ labeled neighbors from the candidate pool. Distance-KNN uses their labels and distances for weighted voting, while the LLM uses the same neighbors as serialized demonstrations, followed by the query features. Retrieval distances are excluded from the LLM prompt. Varying the candidate-pool size while fixing $k$ enables a controlled comparison on identical retrieved examples.}
\label{fig:method}
\vspace{-2mm}
\end{figure*}

\paragraph{Retrieval-Based In-Context Learning with LLMs}
In-context learning (ICL) enables LLMs to adapt to new tasks using a few labeled demonstrations without fine-tuning \citep{brown2020language}. Subsequent studies have examined how models exploit these demonstrations, analyzing the effects of input distributions \citep{min-etal-2022-rethinking}, label mappings \citep{zhao2021,sun-etal-2025-interpret}, and prompt formatting \citep{peng-etal-2025-enhancing}. Demonstration selection has also been widely explored through semantic similarity \citep{liu-etal-2022-makes}, learned retrieval \citep{rubin-etal-2022-learning,liu-etal-2025-unraveling}, and influence estimation \citep{zhang-etal-2025-linear,du-etal-2025-correlation}. In tabular classification, \citet{wu-hou-2025-efficient} propose a training-free retrieval-based approach that prompts LLMs with relevant labeled examples, while InsightTab distills task-specific insights and combines them with selected demonstrations for classification \citep{yuan-etal-2025-summarize}. However, these studies primarily investigate how retrieval strategies or additional information can improve LLM predictions. In contrast, our work compares frozen LLMs with Distance-KNN using identical retrieved neighbors, revealing when LLM inference provides additional predictive value and how these gains depend on neighborhood structure of demonstrations.

\section{Experimental Setup}
\label{sec:setup}

For each test query, a shared retriever extracts the $k$ nearest (most similar) labeled training instances based on feature-space distance. Distance-KNN predicts from their labels and metric distances, whereas the LLM receives their serialized features and labels followed by the query features. By scaling the candidate-pool size under a fixed demonstration budget ($k=8$), we compare their predictive behaviors on identical retrieved evidence (Figure~\ref{fig:method}).

\subsection{Models}
We evaluate four open-weighted LLMs: Qwen3-8B and Qwen3-32B chat models \citep{qwen2025}, alongside instruction-tuned Gemma3-12B and Gemma3-27B \citep{gemma2025}. All generative LLMs are deployed locally via vLLM without quantization. Qwen3-Embedding-4B \citep{qwen3embedding} is employed for text embedding retrieval in \S\ref{sec:complement}. Throughout our experiments, Distance-KNN serves as the reference predictor, using the exact same retrieved neighbors as the LLMs. Full checkpoint identifiers and implementation details are provided in Appendix~\ref{app:modelsettings}.

\subsection{Datasets}
We experiment on 19 public tabular classification datasets from UCI Machine Learning Repository \citep{kelly2023uci} and OpenML \citep{vanschoren2014openml}. These datasets span binary and multiclass tasks across diverse domains, including healthcare, education, finance, industry, and software engineering. Each dataset is partitioned into training, validation, and test splits with the ratio of 8:1:1 via stratified sampling. All predictors share identical splits within each random seed. The training split serves as the retrieval candidate pool. We report macro-averaged accuracy across all datasets to prevent large test sets from dominating aggregate evaluation. Dataset statistics are detailed in Appendix~\ref{app:datasets}. 

\subsection{Protocol}
\label{sec:protocol}
\paragraph{Preprocessing.}
Feature types are determined from the training split. Integer-valued numerical features with at most 20 unique values are treated as categorical. Numerical features are standardized via Z-score normalization, with imputation and standardization statistics fitted strictly on the current candidate pool. 

\paragraph{Candidate pools.}
For each dataset and random seed, we construct a fixed candidate sequence shared by all test queries and predictors. To ensure initial class coverage, we first randomly draw one training example from each class, and then sample remaining examples in proportion to class frequencies. Candidate pools of size $n\in\{16,32,64,128,256,512\}$ are formed as nested prefixes of this sequence, with the full pool comprising the entire training split. For any query $x$, let $N_k(x;P_n)$ denote the $k$ nearest neighbors retrieved from pool $P_n$.

\paragraph{Shared feature-space retrieval and voting.}
Both predictors use the same $k$ nearest neighbors, selected using a mixed-feature distance:
\begin{equation}
d(x,z)=\sqrt{\sum_{j\in\mathcal{N}}(\tilde{x}_j-\tilde{z}_j)^2+\sum_{j\in\mathcal{C}}\mathbf{1}[x_j\neq z_j]},
\end{equation}
where $\mathcal{N}$ and $\mathcal{C}$ denote numerical and categorical features, respectively, and tildes denote standardized values. 
Distance-KNN couples nearest-neighbor retrieval with distance-weighted voting \citep{cover1967,dudani1976}.
Predictions are aggregated via inverse distance weighting. Tie-breaking and near-zero distance handling are detailed in Appendix~\ref{app:aggregation}.

\paragraph{LLM inference.}
The prompt concatenates the task description, feature names, label verbalizers, and the retrieved demonstrations. Demonstrations are serialized as feature key--value pairs with their ground-truth labels and ordered from nearest to farthest, followed by the query features. Predictions are obtained greedily (temperature $0$) from label-token log-probabilities. Crucially, the LLM receives the exact same retrieved neighbors as Distance-KNN, with retrieval distances omitted from the prompt. The prompt template is provided in Appendix~\ref{app:prompt}.

\paragraph{Consensus and proximity.}
Label consensus summarizes neighborhood label homogeneity. Following \citet{klingspohn-etal-2017-applicability}, we define consensus as the majority class fraction among the retrieved neighbors:
\begin{equation}
C(x;P_n)=\frac{1}{k}\max_c\sum_{i\in N_k(x;P_n)}\mathbf{1}[y_i=c].\label{eq:consensus}
\end{equation}
For example, among eight retrieved neighbors, a 4–4 label split yields a consensus of 4/8, whereas a 7–1 split yields 7/8. Unlike Distance-KNN voting, this measure does not weight labels by their retrieval distances.

We quantify neighborhood proximity \citep{klingspohn-etal-2017-applicability} using the mean distance between a test query and its retrieved neighbors ($k=8$):
\begin{equation}\bar d_k(x;P_n)=\frac{1}{k}\sum_{i\in N_k(x;P_n)}d(x,x_i).\label{eq:mean_neighbor_distance}
\end{equation}
For each test query, we compute the mean distance to its retrieved neighbors, with smaller distances indicating closer neighborhoods in feature space. Within each dataset and seed, these distances are ranked across test queries and converted into percentiles to enable comparisons of relative proximity across datasets. Lower percentiles indicate closer neighborhoods, with the 0–20 percentile range representing the closest group. We examine LLM performance by label consensus in \S\ref{sec:rq2} and by its interaction with proximity in \S\ref{sec:rq3}. Details of percentile computation are provided in Appendix~\ref{app:neighborhood}.

\paragraph{Evaluation.}
Accuracy is first averaged over several random seeds per dataset, and then macro-averaged across datasets \citep{gardner2024,wu-hou-2025-efficient}. Mean rank and wins are computed using seed-averaged accuracy. For paired comparisons, we report 95\% bootstrap confidence intervals alongside exact sign-flip tests, with Holm-Bonferroni correction across four LLMs at each pool size \citep{dror-etal-2018-hitchhikers}. Shaded bands show $\pm1$ standard deviation across seeds. 

\begin{figure*}[t]
\centering
\includegraphics[width=0.9\textwidth]{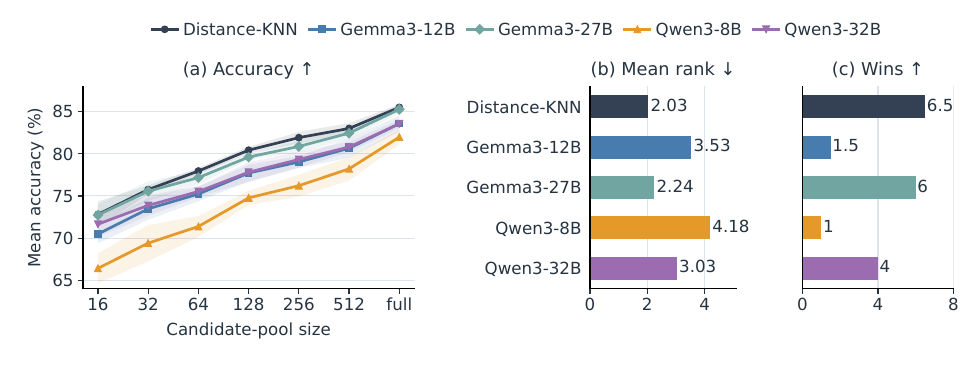}
\vspace{-2mm}
\caption{Performance of Distance-KNN and four LLMs using identical retrieved examples ($k=8$) across 19 tabular prediction datasets with five seeds. (a) Mean accuracy as candidate-pool size increases, with shaded bands indicating $\pm1$ standard deviation across seeds. (b) Mean rank across datasets at full pools (lower is better), with methods ranked by seed-averaged accuracy and ties assigned average ranks. (c) Number of wins at full pools (higher is better), with each dataset contributing one win divided equally among methods tied for the highest accuracy.}
\label{fig:mainresults}
\end{figure*}

\section{Results and Discussion}
\label{sec:results}

\subsection{Do LLMs Outperform Distance-KNN as the Candidate Pool Expands?}
\label{sec:rq1}
\textbf{Larger candidate pools improve all predictors, but Distance-KNN maintains the highest mean accuracy.} 
Figure~\ref{fig:mainresults} summarizes the comparison across seven candidate-pool sizes with the retrieval budget fixed at $k=8$. As shown in Figure~\ref{fig:mainresults}(a), mean accuracy increases consistently for all methods as the candidate pool expands, with Distance-KNN maintaining the highest accuracy at every pool size. Distance-KNN improves from 72.81\% at 16 candidates to 85.50\% at the full pool, while the strongest LLM, Gemma3-27B, follows a similar trajectory, rising from 72.73\% to 85.26\%. These results indicate that larger candidate pools benefit both approaches, but do not yield an aggregate accuracy advantage for LLM inference over distance-weighted voting.

\textbf{Distance-KNN also leads in mean rank and wins, although its advantage over the strongest LLM is marginal.} 
Figure~\ref{fig:mainresults}(b--c) further shows that Distance-KNN attains the lowest mean rank (2.03) and the highest wins (6.5) at the full pool, compared with 2.24 and 6.0 for Gemma3-27B. Paired comparisons in Table~\ref{tab:significance} show that Distance-KNN significantly outperforms three of the four LLMs after Holm correction, further substantiating its effectiveness. These results indicate that LLM inference does not improve mean accuracy over Distance-KNN, motivating a closer examination of how their relative performance varies with label consensus among the retrieved examples. Full-pool dataset-level accuracies are reported in Appendix~\ref{app:accuracy}. 

\begin{table}[t!]
\centering
\setlength{\tabcolsep}{4pt}
\begin{tabular}{lrrr}
\toprule
Model & $\Delta\mathrm{Acc}$ & 95\% CI & $p_{\rm Holm}$\\
\midrule
Gemma3-12B & -1.93 & [-3.31, -0.79] & 0.001 \\
Gemma3-27B & -0.24 & [-1.12, 0.53] & 0.625 \\
Qwen3-8B & -3.53 & [-5.42, -1.84] & $<0.001$ \\
Qwen3-32B & -1.95 & [-3.30, -0.77] & 0.012 \\
\bottomrule
\end{tabular}
\caption{Statistical comparison of full-pool accuracy between Distance-KNN and four LLMs across 19 datasets. Paired differences $\Delta\mathrm{Acc}=\mathrm{Acc}_{\mathrm{LLM}}-\mathrm{Acc}_{\mathrm{KNN}}$ are computed after averaging five seeds. We report 95\% paired percentile-bootstrap confidence intervals over datasets and exact sign-flip $p$-values adjusted for four comparisons using the Holm correction.}
\label{tab:significance}
\end{table}

\paragraph{Takeaway.}
With the same retrieved examples, both LLMs and Distance-KNN benefit from larger candidate pools, yet LLM inference does not surpass distance-weighted voting in mean accuracy at any evaluated pool size.

\subsection{When Does LLM Inference Add Value Beyond Distance-KNN?}
\label{sec:rq2}
\begin{figure}[t]
\centering
\includegraphics[width=0.9\linewidth]{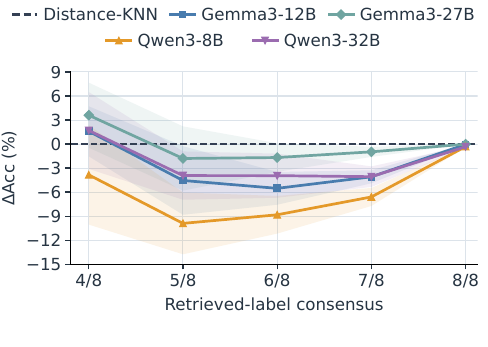}
\vspace{-3mm}
\caption{Accuracy differences between LLMs and Distance-KNN by retrieved-label consensus at full training pools ($k=8$). Consensus is the majority-label fraction among retrieved neighbors. Curves show mean $\Delta\mathrm{Acc}=\mathrm{Acc}_{\mathrm{LLM}}-\mathrm{Acc}_{\mathrm{KNN}}$ across 19 datasets and five seeds. Shaded bands denote $\pm1$ standard deviation.}
\label{fig:consensus}
\end{figure}

To further examine when LLM inference adds value beyond Distance-KNN, we group test queries by label consensus among the shared retrieved examples. Figure~\ref{fig:consensus} reports the mean accuracy differences of LLMs relative to Distance-KNN under the full-pool setting.

\textbf{Low label consensus creates selective opportunities for LLM gains.} 
At the lowest reported consensus level ($4/8$), three of the four LLMs surpass Distance-KNN with positive mean accuracy gaps, led by Gemma3-27B showing the largest improvement of 3.60 percentage points. Qwen3-8B, however, falls behind Distance-KNN, indicating that low consensus does not benefit all models equally. The override analysis in Appendix~\ref{app:override} shows that beneficial corrections outweigh harmful overrides for the three improving models. These results suggest that some LLMs can leverage information in mixed-label demonstrations beyond what is captured by distance-weighted voting.

\textbf{High label consensus favors Distance-KNN.} 
Across consensus levels of $5/8$--$7/8$, Distance-KNN consistently outperforms all evaluated LLMs. The most pronounced deficit occurs for Qwen3-8B, which trails Distance-KNN by 9.91 percentage points at $5/8$ consensus, while Gemma3-27B remains the closest competitor throughout this range. As detailed in Appendix~\ref{app:override}, harmful overrides exceed beneficial corrections across all models in these groups. This suggests that LLM inference can disrupt useful local evidence when the retrieved examples exhibit a high label consensus. At full label consensus ($8/8$), both types of overrides become rare, and the accuracy gaps approach zero.

\paragraph{Takeaway.}
The benefits of LLM inference are concentrated in low label consensus neighborhoods and depend on the model, whereas higher label consensus generally favors distance-weighted voting.

\subsection{How Do Label Consensus and Feature Proximity Jointly Shape LLM Gains?}
\label{sec:rq3}
\begin{figure}[t]
\centering
\includegraphics[width=0.9\linewidth]{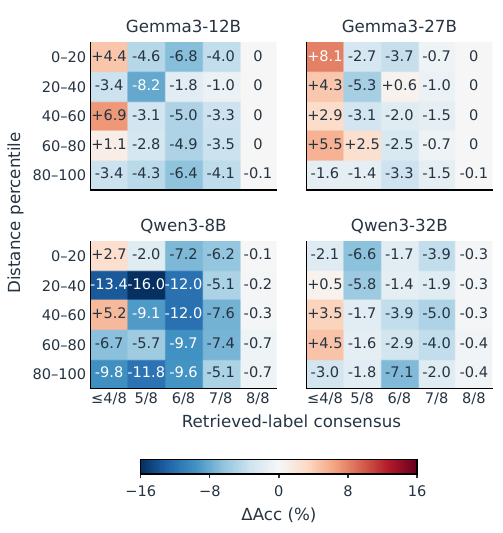}
\caption{Accuracy differences between LLMs and Distance-KNN across retrieved-label consensus ($k=8$) and feature proximity at full training pools. Cells show mean $\Delta\mathrm{Acc}=\mathrm{Acc}_{\mathrm{LLM}}-\mathrm{Acc}_{\mathrm{KNN}}$ across datasets. Columns indicate label consensus, while rows group test queries by distance percentile based on mean neighbor distance per dataset and seed (lower values indicate closer neighborhoods).}
\label{fig:heatmap}
\end{figure}

To examine whether feature proximity further distinguishes LLM gains under different levels of label consensus, Figure~\ref{fig:heatmap} shows how the accuracy gaps between LLMs and Distance-KNN vary across combinations of label consensus and feature proximity. Distance percentiles are computed from the mean distance to the eight shared neighbors within each dataset and seed, with lower values indicating closer neighborhoods (Appendix~\ref{app:neighborhood}).

\textbf{Under low label consensus, feature proximity further distinguishes LLM gains and losses.} 
At $\leq 4/8$ consensus, three LLMs outperform Distance-KNN in the closest stratum (0--20 percentile), with Gemma3-27B reaching a gain of 8.1\%. Notably, all four LLMs achieve positive gaps at the intermediate 40--60 percentile, showing that the advantages are not restricted to the closest neighborhoods. Conversely, all four models fall behind Distance-KNN at the most distant 80--100 percentile, with accuracy gaps ranging from $-9.8\%$ to $-1.6\%$. These results indicate that feature proximity helps characterize LLM gains under label disagreement, although closer neighbors do not consistently yield larger advantages.

\begin{figure*}[t]
\centering
\includegraphics[width=0.9\textwidth]{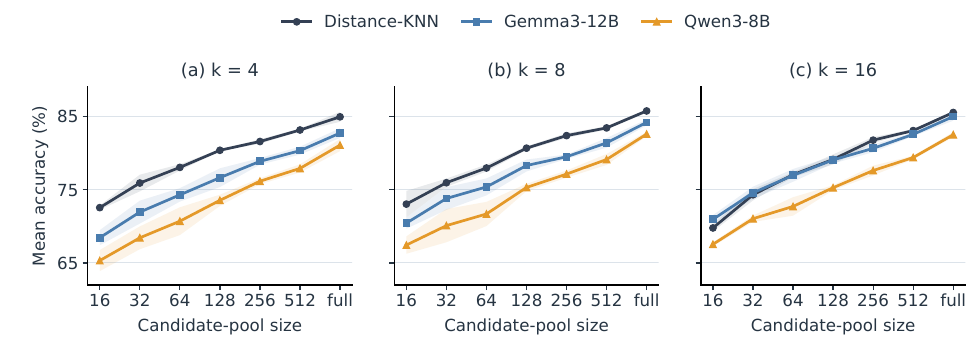}
\vspace{-5mm}
\caption{Effect of the number of demonstrations on classification accuracy across candidate-pool sizes. We compare (a) $k=4$, (b) $k=8$, and (c) $k=16$, with Distance-KNN and LLMs using the same retrieved neighbors at each $k$. Curves show mean accuracy across 19 datasets and three seeds. Shaded bands denote $\pm1$ standard deviation.}
\label{fig:demonstration}
\vspace{-2mm}
\end{figure*}

\textbf{LLM gains largely disappear at higher label consensus, regardless of proximity.} 
At $5/8$--$7/8$ consensus, Distance-KNN outperforms the LLMs in 58 of the 60 model--proximity combinations, including nearly all cases involving the closest neighbors. The only two positive gaps occur for Gemma3-27B, indicating that proximity alone is insufficient to recover a consistent LLM advantage when label consensus increases. At full consensus ($8/8$), the absolute accuracy gaps shrink to at most 0.7\% across all models and proximity strata. 
Together, these findings suggest that feature proximity is most informative for distinguishing LLM gains under low label consensus, while higher consensus consistently limits such gains across proximity levels.

\paragraph{Takeaway.} 
Feature proximity further differentiates LLM gains within low-consensus neighborhoods. LLMs benefit from selected close and intermediate neighborhoods but consistently fall behind Distance-KNN in the most distant ones. These gains remain model-dependent and largely disappear as label consensus increases.

\section{Complementary Analyses}
\label{sec:complement}
We conduct four complementary analyses to examine how the number of demonstrations, prompt formulation, retrieval representation, and label-balanced selection affect the comparison between LLM inference and distance-weighted voting. All experiments cover 19 datasets and three seeds using Gemma3-12B and Qwen3-8B. The first two analyses retain feature-based nearest-neighbor retrieval, while the latter two examine alternative retrieval representations and label-balanced selection, respectively.

\paragraph{Number of demonstrations.}
We vary the number of demonstrations $k \in \{4, 8, 16\}$ to test whether the findings in \S\ref{sec:rq1} hold beyond the default setting of $k=8$. Figure~\ref{fig:demonstration} compares accuracy across all candidate-pool sizes, using the same retrieved examples for each LLM and Distance-KNN. Results show that mean accuracy increases with candidate-pool size for all predictors across three demonstration budgets, consistent with the results in Figure~\ref{fig:mainresults}. Distance-KNN also maintains the highest mean accuracy in 19 of the 21 pool--$k$ settings, with Gemma3-12B slightly ahead at the two smallest pools when $k=16$. Increasing $k$ benefits Gemma3-12B more consistently than Qwen3-8B, while Distance-KNN shows no further improvement beyond eight neighbors. Overall, the main findings in Section~\ref{sec:rq1} remain largely consistent across demonstration budgets, with limited exceptions at small candidate pools. Additional results are provided in Appendix~\ref{app:robustness}.

\begin{figure*}[t]
\centering
\includegraphics[width=0.88\textwidth]{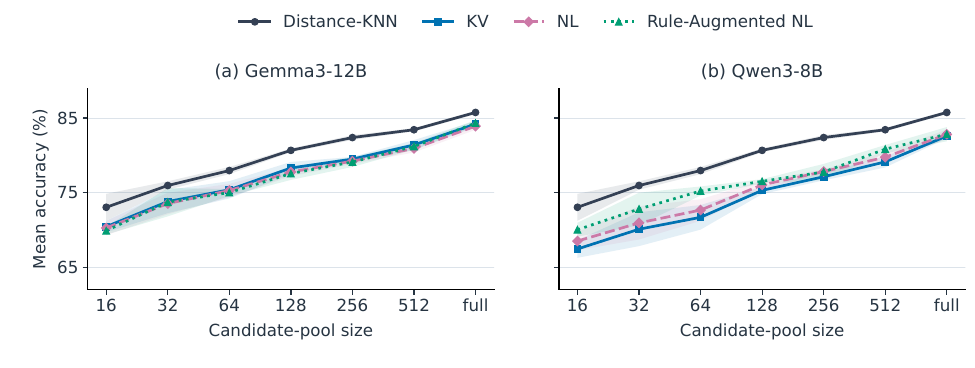}
\vspace{-5mm}
\caption{Effect of prompt formulation on classification accuracy for (a) Gemma3-12B and (b) Qwen3-8B across KV, NL and Rule-Augmented NL using identical retrieved examples ($k=8$). Curves show mean accuracy across 19 datasets and three seeds as candidate-pool size increases. Shaded bands denote $\pm1$ standard deviation.}
\label{fig:presentation}
\end{figure*}

\paragraph{Prompt formulation.}
To examine whether the default prompt formulation limits LLM performance, we compare three formulations: key--value serialization (KV), natural-language serialization (NL), and rule-augmented natural-language serialization (Rule-Augmented NL). Our main experiments use KV, which presents features as key--value pairs, whereas NL expresses them as natural-language statements. Rule-augmented NL further supplements NL with training-derived classification rules using a procedure adapted from InsightTab \citep{yuan-etal-2025-summarize}. All three formulations use the same eight retrieved demonstrations and their order, with additional rules supplied only in Rule-Augmented NL. 

Figure~\ref{fig:presentation} shows that prompt formulation affects the two LLMs differently, while Distance-KNN maintains the highest mean accuracy across all seven candidate-pool sizes. Gemma3-12B generally performs best with KV, with only minor differences among formulations. In contrast, Qwen3-8B consistently benefits from NL and Rule-Augmented NL over KV, with the latter achieving a notable accuracy gain of 3.53\% at 64 candidates. These observations indicate that prompt formulation can improve LLM predictions, but the performance gap identified in \S\ref{sec:rq1} is not solely attributable to the default prompt formulation. Prompt templates and rule construction are detailed in Appendix~\ref{app:presentation}.

\paragraph{Retrieval representation.}
We compare the default feature-space retrieval (as defined in \S\ref{sec:protocol}) with text-embedding retrieval, which represents serialized tabular instances using Qwen3-Embedding-4B. Table~\ref{tab:retrieval_summary} reports the full-pool comparison, with results across all candidate-pool sizes in Appendix Table~\ref{tab:embedding}. Feature-space retrieval achieves higher mean accuracy for all three predictors at every evaluated pool size. At the full pool, switching to text-embedding retrieval reduces accuracy by approximately 16 percentage points for both Distance-KNN and Gemma3-12B, and by 24.20 points for Qwen3-8B. Despite these changes, Distance-KNN maintains the highest mean accuracy under both retrieval representations. These results extend \S\ref{sec:rq1} by showing that retrieval representation strongly influences prediction quality, but changing the retrieval space does not lead to a consistent advantage for LLM inference.

\begin{table}[t]
\centering
\setlength{\tabcolsep}{3pt}
\begin{tabular}{@{}lccc@{}}
\toprule
Model & Feat-Space & Text-Embed & $\Delta\mathrm{Acc}$\\
\midrule
Distance-KNN & \textbf{85.75} & 69.89 & +15.86\\
Gemma3-12B & \textbf{84.19} & 68.51 & +15.68\\
Qwen3-8B & \textbf{82.59} & 58.39 & +24.20\\
\bottomrule
\end{tabular}
\caption{Effect of retrieval representation on classification accuracy for Distance-KNN and two LLMs at full training pools. Feature-space retrieval is compared against text-embedding retrieval ($k=8$) across 19 datasets and three seeds.}
\label{tab:retrieval_summary}
\vspace{-3mm}
\end{table}

\paragraph{Label-balanced demonstration selection.}
To complement our analyses of label consensus and feature proximity in \S\ref{sec:rq2} and \S\ref{sec:rq3}, we compare the nearest eight demonstrations with eight label-balanced examples selected from the nearest 32 candidates at full training pools. Table~\ref{tab:balance_summary} shows that this strategy reduces mean accuracy by 7.49 and 10.58 percentage points for Gemma3-12B and Qwen3-8B, respectively, with declines on all 19 datasets. Unlike naturally occurring low-consensus neighborhoods examined earlier, this procedure increases label diversity by replacing some nearby examples with more distant ones, changing both label consensus and feature proximity. These results indicate that enforcing label balance does not improve LLM accuracy compared with nearest-neighbor selection.

\begin{table}[t]
\centering
\setlength{\tabcolsep}{3pt}
\begin{tabular}{@{}lccc@{}}
\toprule
Model & Nearest & Label-Balanced & $\Delta\mathrm{Acc}$\\
\midrule
Gemma3-12B & \textbf{84.19} & 76.70 & -7.49\\
Qwen3-8B   & \textbf{82.59} & 72.01 & -10.58\\
\bottomrule
\end{tabular}
\caption{Effect of demonstration selection on classification accuracy for two LLMs at full training pools. The eight nearest neighbors are compared with eight label-balanced examples selected from the 32 nearest candidates ($k=8$) across 19 datasets and three seeds.}
\label{tab:balance_summary}
\vspace{-3mm}
\end{table}

\section{Conclusion}
We investigate whether and when LLMs add predictive value beyond distance-weighted voting on the same retrieved tabular examples. Across 19 datasets and expanding candidate pools, Distance-KNN achieves the highest mean accuracy at every pool size, although its margin over the strongest LLM is small. Some LLMs outperform voting under low label consensus, with gains varying by model and feature proximity. As label consensus increases, Distance-KNN generally retains the advantage, and low-consensus gains do not extend to the most distant neighborhoods. Complementary analyses indicate that changing the number of demonstrations, prompt formulation, retrieval representation, or label-balanced selection does not substantially alter these findings. Our work highlights the importance of comparing LLMs against traditional classifiers on the same retrieved examples and motivates further work on identifying the neighborhoods in which LLMs are most useful.

\section*{Limitations}
Our study focuses on the predictive value of general-purpose LLMs given the same retrieved neighbors as Distance-KNN. Although this design controls demonstration selection, it does not evaluate independently optimized end-to-end systems or task-specific model adaptation. Moreover, while our evaluation spans 19 diverse datasets, validating whether these behavioral patterns generalize to other emerging tabular benchmarks or highly domain-specific datasets remains a promising direction for future work.

\section*{Ethical Considerations}
Our study uses publicly available benchmark datasets and existing pretrained language models, involving no new data collection or human annotation. Our experiments aim to provide empirical insights into retrieval-based tabular prediction rather than establish the suitability of these methods for real-world deployment. Downstream applications in sensitive domains would require additional evaluation of reliability and fairness.

\bibliography{main}

\appendix
\raggedbottom

\setcounter{topnumber}{3}
\setcounter{bottomnumber}{2}
\setcounter{totalnumber}{5}
\setcounter{dbltopnumber}{2}
\setlength{\textfloatsep}{10pt plus 2pt minus 2pt}
\setlength{\floatsep}{8pt plus 2pt minus 2pt}
\setlength{\intextsep}{8pt plus 2pt minus 2pt}
\setlength{\dbltextfloatsep}{12pt plus 2pt minus 2pt}
\setlength{\dblfloatsep}{10pt plus 2pt minus 2pt}
\renewcommand{\topfraction}{0.90}
\renewcommand{\dbltopfraction}{0.90}
\renewcommand{\textfraction}{0.07}
\renewcommand{\floatpagefraction}{0.70}
\renewcommand{\dblfloatpagefraction}{0.70}

\section{Evaluation Protocol and Reproducibility}
\label{app:evaluation}

\subsection{Models and inference}
\label{app:modelsettings}
\paragraph{Checkpoints.}
We use Qwen3-8B and Qwen3-32B post-trained chat models \citep{qwen2025}, and gemma-3-12b-it and gemma-3-27b-it instruction-tuned models \citep{gemma2025}. All LLM parameters remain frozen. Models run locally through vLLM \citep{kwon2023efficient} without quantization: Qwen3-8B and Gemma3-12B each use one NVIDIA GeForce RTX 4090, while Qwen3-32B and Gemma3-27B each use one NVIDIA RTX PRO 6000 Blackwell.

\paragraph{Label scoring.}
Classes are mapped to numeric identifiers $1,\ldots,C$, where $C \le 10$ is the number of classes. Qwen3 uses \texttt{enable\_thinking=False}.
Each request returns at most one output token and the top 20 first-token log-probabilities. After stripping token whitespace, we select the highest-scoring valid numeric label; labels absent from the returned list receive a score of $-\infty$.

\paragraph{Embedding retrieval.}
Qwen/Qwen3-Embedding-4B runs through Hugging Face Transformers in FP16 without quantization on a single NVIDIA GeForce RTX 4090. Rows are serialized as compact feature--value JSON, excluding target labels. Query inputs use the \texttt{Instruct: ...} and \texttt{Query: ...} format with the instruction ``Given a tabular record, retrieve training records with similar feature values for classification.'' Candidate rows receive no instruction prefix. Last-token pooling and $L_2$ normalization produce 2,560-dimensional embeddings. Retrieval uses cosine similarity, and embedding-KNN uses inverse-distance voting with $d_{\mathrm{emb}}=1-\cos(e_x,e_i)$.

\subsection{Datasets and analysis cohorts}
\label{app:datasets}
Table~\ref{tab:datasets} lists dataset characteristics and exact split sizes under the partitioning protocol described in Section~\ref{sec:setup}. Feature counts refer to the processed inputs; dataset identifiers and target columns are specified in the released configuration. The main comparison covers 19 datasets, five seeds, seven candidate pools, and five predictors. The four complementary analyses use the same 19 datasets and three seeds; label-balanced retrieval uses full pools, and the other three analyses use all seven pool sizes.

\begin{table*}[t]
\centering
\small
\setlength{\tabcolsep}{12pt}
\begin{tabular}{lrrrl}
\toprule
Dataset & Rows & Features & Classes & Train / validation / test\\
\midrule
AI4I Predictive Maintenance & 10,000 & 6 & 2 & 8,000 / 1,000 / 1,000\\
Car & 1,728 & 6 & 4 & 1,382 / 173 / 173\\
Cardiotocography & 2,126 & 21 & 3 & 1,700 / 213 / 213\\
Credit G & 1,000 & 20 & 2 & 800 / 100 / 100\\
Diabetes & 768 & 8 & 2 & 614 / 77 / 77\\
Dry Bean & 13,611 & 16 & 7 & 10,888 / 1,361 / 1,362\\
Glioma Grading & 839 & 23 & 2 & 671 / 84 / 84\\
Iranian Churn & 3,150 & 13 & 2 & 2,520 / 315 / 315\\
KC1 & 2,109 & 21 & 2 & 1,687 / 211 / 211\\
MAGIC Gamma & 19,020 & 10 & 2 & 15,216 / 1,902 / 1,902\\
Online Shoppers & 12,330 & 17 & 2 & 9,864 / 1,233 / 1,233\\
PC3 & 1,563 & 37 & 2 & 1,250 / 156 / 157\\
PC4 & 1,458 & 37 & 2 & 1,166 / 146 / 146\\
Phishing Websites & 11,055 & 30 & 2 & 8,844 / 1,105 / 1,106\\
QSAR Biodegradation & 1,055 & 41 & 2 & 844 / 105 / 106\\
Spambase & 4,601 & 57 & 2 & 3,680 / 460 / 461\\
Steel Plates Fault & 1,941 & 27 & 7 & 1,552 / 194 / 195\\
Student Dropout & 4,424 & 36 & 3 & 3,539 / 442 / 443\\
Vehicle & 846 & 18 & 4 & 676 / 85 / 85\\
\bottomrule
\end{tabular}
\caption{Characteristics and train/validation/test split sizes of the 19 tabular classification datasets. Splits are stratified by class in an 8:1:1 ratio, and feature counts refer to the processed inputs.}
\label{tab:datasets}
\end{table*}

\subsection{Implementation and statistics}
\label{app:aggregation}

\paragraph{Preprocessing details.}
Identifier-like and constant columns are removed by the dataset loader. Missing numerical values are imputed using candidate-pool medians; remaining categorical missing values use a pool-derived mode. Categorical values are converted to strings before overlap comparison.

\paragraph{Voting details.}
If any retrieved distance is at most $10^{-12}$ (indicating ``zero-distance'' neighbors that almost exactly coincide with the test sample), only these identical neighbors are trusted to vote with equal weights, while all other retrieved neighbors are excluded from voting. Class-score ties are resolved by the first retrieved neighbor whose label belongs to a tied class.

\paragraph{Statistical details.}
Percentile-bootstrap intervals use 10,000 paired dataset resamples and the 2.5th and 97.5th percentiles. Reported sign-flip tests enumerate all $2^{19}$ sign assignments. Holm correction applies across the four LLM comparisons within each metric and candidate-pool size. Bootstrap intervals are pointwise, not simultaneous confidence bands. Seed standard deviations use the sample estimator with denominator $S-1$.

\subsection{Classification prompt}
\label{app:prompt}
\label{app:reproducibility}
The main classification prompt is shown below. Label scoring follows Appendix~\ref{app:modelsettings}.

\begin{promptbox}{Retrieved-Example-Based Classification}
You are a helpful data analyst.\par
Task description:\par
\{task\_description\}\par
\smallskip
Features:\par
\{feature\_names\}\par
\smallskip
Label mapping:\par
\{label\_mapping\}\par
\smallskip
The following are the top-k most similar training instances retrieved
for the target instance. They are ordered from the most similar to the
least similar.\par
Use these retrieved examples to infer the label of the target
instance.\par
\smallskip
CRITICAL: Your response must be EXACTLY one number
from \{candidate\_labels\}.\par
Output ONLY the number, nothing else.
No explanations, no reasoning, no tags, no extra text.\par
\smallskip
Labeled instances:\par
\{retrieved\_examples\}\par
\smallskip
Target instance:\par
\{target\_instance\}\par
\smallskip
Your answer (one number only):
\end{promptbox}

\section{Supplementary Results for RQ1--RQ3}
\label{app:fullresults}

\subsection{RQ1: Dataset-level accuracy}
\label{app:accuracy}
Table~\ref{tab:datasetacc} reports full-pool dataset-level accuracy.

\begin{table*}[!t]
\centering
\small
\setlength{\tabcolsep}{5pt}
\begin{tabular*}{\textwidth}{@{\extracolsep{\fill}}lrrrrr@{}}
\toprule
Dataset & Distance-KNN & Gemma3-12B & Gemma3-27B & Qwen3-8B & Qwen3-32B\\
\midrule
AI4I Maintenance & \textbf{97.58} & 97.56 & 97.32 & 96.78 & 97.04 \\
Car & 90.17 & 89.25 & \textbf{93.06} & 89.83 & 90.87 \\
Cardiotocography & 91.92 & 91.27 & 92.02 & 90.42 & \textbf{93.52} \\
Credit G & \textbf{73.60} & \textbf{73.60} & 73.00 & 73.20 & 71.40 \\
Diabetes & \textbf{74.29} & 63.90 & 68.31 & 63.12 & 69.35 \\
Dry Bean & 92.29 & 91.78 & 92.13 & 92.20 & \textbf{92.35} \\
Glioma Grading & 84.52 & 85.24 & \textbf{86.19} & 84.29 & 83.33 \\
Iranian Churn & 94.35 & 93.71 & \textbf{94.79} & 91.75 & 94.10 \\
KC1 & \textbf{86.82} & 80.57 & 85.59 & 75.26 & 80.38 \\
MAGIC Gamma & 84.54 & 84.22 & \textbf{84.71} & 84.64 & 82.64 \\
Online Shoppers & \textbf{87.98} & 82.32 & 86.08 & 78.02 & 78.98 \\
PC3 & \textbf{89.30} & 85.35 & 88.54 & 85.35 & 86.24 \\
PC4 & \textbf{88.49} & 84.11 & 87.40 & 82.88 & 84.79 \\
Phishing Websites & 96.93 & \textbf{97.00} & 96.73 & 86.71 & 96.93 \\
QSAR Biodegradation & 85.28 & 84.34 & 85.28 & \textbf{85.66} & 84.72 \\
Spambase & 91.28 & 90.54 & 90.11 & 87.20 & \textbf{91.63} \\
Steel Plates Fault & 75.18 & 73.85 & 75.38 & 74.46 & \textbf{75.59} \\
Student Dropout & 70.74 & 70.56 & \textbf{73.18} & 68.89 & 64.15 \\
Vehicle & 69.18 & 68.71 & \textbf{70.12} & 66.82 & 69.41 \\

\bottomrule
\end{tabular*}
\caption{Full-pool classification accuracy (\%) of Distance-KNN and four LLMs on 19 datasets using shared retrieved neighbors ($k=8$). Each entry averages five seeds. Bold marks the highest displayed accuracy within each dataset, including ties.}
\label{tab:datasetacc}
\end{table*}

\subsection{RQ2: Complementary predictions and overrides}
\label{app:override}
\begin{figure}[t]
\centering
\includegraphics[width=\linewidth]{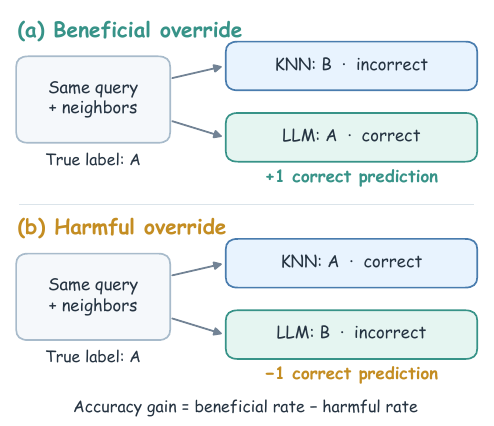}
\caption{Schematic illustration of beneficial and harmful overrides on shared queries and retrieved neighbors. Beneficial overrides occur when an LLM predicts correctly and Distance-KNN predicts incorrectly, while harmful overrides reverse these outcomes.}
\label{fig:override}
\end{figure}

Figure~\ref{fig:override} illustrates the beneficial and harmful cases defined in Section~\ref{sec:rq2}. 

\begin{figure}[!htbp]
\centering
\includegraphics[width=0.9\linewidth]{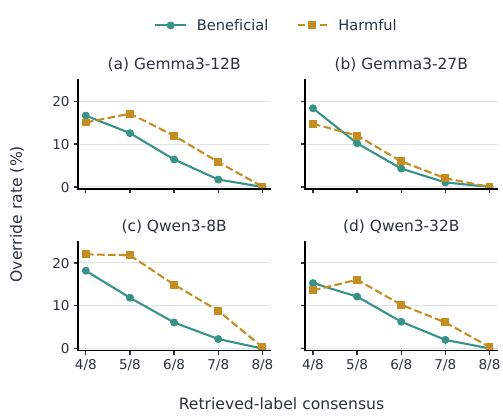}
\caption{Beneficial ($B$) and harmful ($H$) override rates (\%) by retrieved-label consensus at full training pools ($k=8$). Each rate uses all queries within its dataset--seed--consensus group as the denominator. Rates are averaged equally over non-empty dataset groups within each seed and then over five seeds. As shown in Figure~\ref{fig:consensus}, the mean accuracy difference satisfies $\text{Acc}_{\text{LLM}} - \text{Acc}_{\text{KNN}} = B - H$.}
\label{fig:overriderates}
\end{figure}

\paragraph{Aggregation and empty groups.}
At each consensus level, we compute beneficial and harmful override rates separately for each dataset and seed. The denominator is the total number of queries at that consensus level. Within each seed, we average these rates equally across datasets with non-empty groups, and then average the resulting values across the five seeds. Empty groups are excluded rather than assigned a rate of zero. Figures~\ref{fig:overriderates} and~\ref{fig:consensus} use the same query pairing and aggregation procedure.

The displayed consensus levels are $4/8$--$8/8$. At $4/8$, Glioma Grading and PC3 each have one empty seed-level group, yielding 18 or 19 datasets per seed. All groups contain 19 datasets at $5/8$--$8/8$. Empty groups are omitted consistently across models, rather than assigned zero.

\paragraph{Dataset-level rates.}
Table~\ref{tab:overridedatasets} reports full-pool rates without consensus conditioning. Each entry averages five seeds; before rounding, beneficial minus harmful equals the corresponding LLM--KNN accuracy difference in Table~\ref{tab:datasetacc}.

\begin{table*}[t]
\centering
\small
\setlength{\tabcolsep}{5pt}
\begin{tabular}{@{}lrrrrrrrr@{}}
\toprule
& \multicolumn{2}{c}{Gemma3-12B}
& \multicolumn{2}{c}{Gemma3-27B}
& \multicolumn{2}{c}{Qwen3-8B}
& \multicolumn{2}{c}{Qwen3-32B}\\
\cmidrule(lr){2-3}\cmidrule(lr){4-5}
\cmidrule(lr){6-7}\cmidrule(lr){8-9}
Dataset & B & H & B & H & B & H & B & H\\
\midrule
AI4I Maintenance & 0.58 & 0.60 & 0.44 & 0.70 & 0.58 & 1.38 & 0.66 & 1.20\\
Car & 5.66 & 6.59 & 5.55 & 2.66 & 4.28 & 4.62 & 5.55 & 4.86\\
Cardiotocography & 1.69 & 2.35 & 1.60 & 1.50 & 1.41 & 2.91 & 2.91 & 1.31\\
Credit G & 4.80 & 4.80 & 5.00 & 5.60 & 4.80 & 5.20 & 5.40 & 7.60\\
Diabetes & 11.17 & 21.56 & 8.57 & 14.55 & 12.21 & 23.38 & 7.79 & 12.73\\
Dry Bean & 0.84 & 1.35 & 0.57 & 0.73 & 0.75 & 0.84 & 0.69 & 0.63\\
Glioma Grading & 4.29 & 3.57 & 4.05 & 2.38 & 5.24 & 5.48 & 4.05 & 5.24\\
Iranian Churn & 1.52 & 2.16 & 1.59 & 1.14 & 1.27 & 3.87 & 1.84 & 2.10\\
KC1 & 4.08 & 10.33 & 1.80 & 3.03 & 6.16 & 17.73 & 3.98 & 10.43\\
MAGIC Gamma & 3.55 & 3.88 & 3.50 & 3.33 & 2.80 & 2.70 & 3.05 & 4.95\\
Online Shoppers & 3.88 & 9.54 & 3.26 & 5.16 & 5.27 & 15.23 & 5.03 & 14.03\\
PC3 & 4.97 & 8.92 & 4.08 & 4.84 & 4.33 & 8.28 & 3.95 & 7.01\\
PC4 & 4.66 & 9.04 & 3.70 & 4.79 & 5.48 & 11.10 & 4.38 & 8.08\\
Phishing Websites & 1.01 & 0.94 & 0.67 & 0.87 & 0.99 & 11.21 & 0.90 & 0.90\\
QSAR Biodegradation & 2.26 & 3.21 & 2.45 & 2.45 & 2.08 & 1.70 & 2.26 & 2.83\\
Spambase & 2.52 & 3.25 & 2.52 & 3.69 & 2.56 & 6.64 & 2.34 & 2.00\\
Steel Plates Fault & 3.79 & 5.13 & 2.46 & 2.26 & 2.15 & 2.87 & 3.28 & 2.87\\
Student Dropout & 5.46 & 5.64 & 5.15 & 2.71 & 3.88 & 5.73 & 8.89 & 15.49\\
Vehicle & 3.76 & 4.24 & 3.06 & 2.12 & 2.82 & 5.18 & 2.82 & 2.59\\
\bottomrule
\end{tabular}
\caption{Full-pool beneficial ($B$) and harmful ($H$) override rates (\%) for four LLMs relative to Distance-KNN across 19 datasets ($k=8$). Within each dataset, rates use all test queries as the denominator for each seed and are averaged over five seeds. Note that $\text{Acc}_{\text{LLM}} - \text{Acc}_{\text{KNN}} = B - H$ (before rounding).}
\label{tab:overridedatasets}
\end{table*}

\subsection{RQ3: Proximity and label composition}
\label{app:geometry}
\label{app:neighborhood}

We measure proximity using the mean distance from each test query $x$ to its eight retrieved neighbors under the full-pool setting, denoted by $\bar d_8(x;P_{\mathrm{full}})$. Because raw distances may have different scales across datasets, we convert them into relative distance percentiles within each dataset and seed.

Specifically, we rank the test queries by their mean neighbor distance in ascending order. The distance percentile is
\begin{equation}
p_{d,s}(x) = 100 \times \frac{ \operatorname{rank}_{\mathrm{avg}} \bigl(\bar d_8(x;P_{\mathrm{full}})\bigr) }{ |\mathcal{Q}_{d,s}| },
\label{eq:locality_percentile}
\end{equation}
where $\mathcal{Q}_{d,s}$ is the set of test queries for dataset $d$ and seed $s$. Ranks start at 1, and tied distances receive their average rank. A lower percentile indicates a relatively closer neighborhood within the same dataset and seed; it does not measure absolute proximity across datasets.

These percentiles are computed from shared neighbor distances, so they remain identical across predictors. They require no ground-truth query labels and serve only as a post-hoc diagnostic, without altering retrieval or prediction. For the heatmap, we group percentiles into five equal-width bands: $(0,20]$, $(20,40]$, $(40,60]$, $(60,80]$, and $(80,100]$. These bands need not contain equal numbers of queries because of tied distances.

\begin{figure}[!htbp]
\centering
\includegraphics[width=0.8\linewidth]{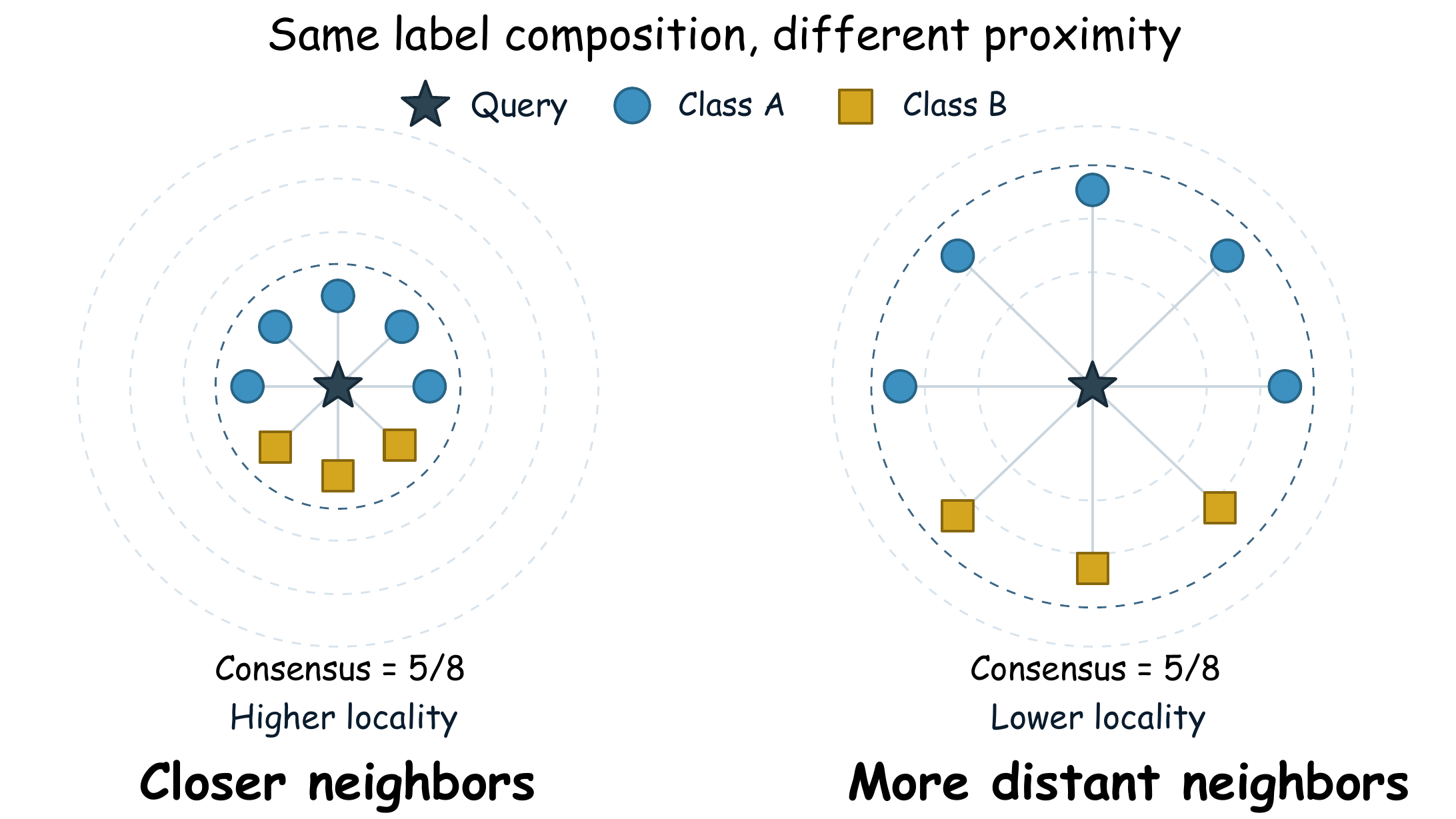}
\caption{Schematic comparison of label consensus and feature proximity. The two neighborhoods have the same label composition, with five of eight neighbors in the majority class, but differ in their distances to the query.}
\label{fig:geometry}
\end{figure}

Figure~\ref{fig:geometry} illustrates how neighborhoods with the same label consensus can differ in proximity.

\section{Robustness and Complementary Analyses}
\label{app:complementary}

\subsection{Demonstration count}
\label{app:robustness}
Figure~\ref{fig:demonstration} compares $k=4,8,16$ on 19 datasets and three seeds. Full-pool Distance-KNN accuracy rises by 0.80 percentage points from $k=4$ to $8$, then falls by 0.21 at $k=16$. Gemma3-12B leads at the two smallest pools with $k=16$, while Distance-KNN leads at full pools for every tested budget.

\subsection{Prompt formulation}
\label{app:presentation}
\paragraph{Conditions.}
KV uses feature key--value pairs; NL uses statements of the form ``the \{feature\} is \{value\}''; Rule-Augmented NL augments NL with rules from labeled candidate-pool data. All conditions share the same eight retrieved examples, labels, and order, and use three seeds.

\paragraph{Rule construction.}
For Rule-Augmented NL, we adapt the Summarize--Exemplify--Reflect rule-distillation procedure of InsightTab \citep{yuan-etal-2025-summarize}. For each dataset, seed, and candidate pool, a class-balanced subset of up to 128 samples trains an auxiliary XGBoost classifier (100 trees, maximum depth 6, learning rate 0.3, exact tree method). The first tree's leaves define sample groups. Predictive entropy guides the selection of eight low-entropy examples and a high-entropy subset containing approximately half of the selected samples, with selection aiming to balance classes.

The classification checkpoint (google/gemma-3-12b-it or Qwen/Qwen3-8B) also summarizes and merges group-wise rules. Using these rules and the low-entropy examples, it predicts the high-entropy examples; errors and their correct labels support one additional round of rule generation. XGBoost assists rule construction, while the LLM makes the final test predictions.

\paragraph{Validation and reuse.}
Rules are JSON objects with one to three feature conditions joined by AND and a predicted class. Generated, merged, and additional rules are evaluated on the entire current candidate pool. Acceptance requires support $\geq3$, coverage $\geq0.01$, precision $\geq0.70$, and a 95\% Wilson lower bound $\geq0.40$; this empirical lower bound replaces the rule confidence. Unparseable outputs produce no rules, invalid rules are discarded, and an empty retained set gives empty rule arrays in the prediction template.

Construction, validation, and refinement use training features and labels only. Rules are cached per dataset, seed, candidate pool, and model configuration, then reused across test queries. The construction examples do not replace the eight retrieved demonstrations. Rule-generation requests use temperature $0$ and at most 512 output tokens; input sections for summarization, merging, and reflection are limited to 24,000 characters, with excess content omitted or truncated. Final classification uses temperature $0$ and first-token label log-probabilities. LLM parameters remain frozen; only the auxiliary XGBoost model is trained.

\paragraph{Prediction template.}
The merged and additional rules below are validated JSON arrays. The labeled examples are the same eight rows used by Distance-KNN.

\begin{promptbox}{Rule-Augmented NL: Classification Prompt}
You are performing tabular classification using distilled
training-data insights.\par
\smallskip

Task description: \{task\_description\}\par
\smallskip

Label mapping:\par
\{label\_mapping\}\par
\smallskip

Useful patterns for the task:\par
\{merged\_rules\}\par
\smallskip

Additional patterns learned from difficult misclassified
examples:\par
\{additional\_rules\}\par
\smallskip

[FEW-SHOT EXAMPLES START]\par
\{eight\_retrieved\_labeled\_examples\}\par
[FEW-SHOT EXAMPLES END]\par
\smallskip

[CURRENT QUESTION START]\par
\{query\_features\_in\_natural\_language\}\par
[CURRENT QUESTION END]\par
\smallskip

Your response must be EXACTLY one number from
\{valid\_label\_ids\}.\par
Output ONLY the number, nothing else:
\end{promptbox}

\subsection{Feature-based versus embedding retrieval}
\label{app:embdatasets}
Table~\ref{tab:embedding} extends Table~\ref{tab:retrieval_summary} to all seven candidate pools. Values average the same 19 datasets and three seeds, using the $k=8$ feature-based results and the embedding results with matched coverage. Feature-based retrieval has higher accuracy in all 21 predictor-by-pool comparisons.

\begin{table*}[t]
\centering
\captionsetup{skip=3pt}
\small
\setlength{\tabcolsep}{3pt}
\begin{tabular*}{\textwidth}{@{\extracolsep{\fill}}lrrrrrr@{}}
\toprule
& \multicolumn{3}{c}{Feature-space retrieval} & \multicolumn{3}{c}{Text-embedding retrieval}\\
\cmidrule(lr){2-4}\cmidrule(lr){5-7}
Pool & \shortstack{Distance-\\KNN} & Gemma3-12B & Qwen3-8B & \shortstack{Distance-\\KNN} & Gemma3-12B & Qwen3-8B\\
\midrule
16 & 73.03 & 70.44 & 67.45 & 67.79 & 67.04 & 60.12\\
32 & 75.95 & 73.80 & 70.10 & 67.46 & 66.43 & 60.80\\
64 & 77.96 & 75.38 & 71.69 & 68.07 & 66.86 & 60.14\\
128 & 80.67 & 78.30 & 75.31 & 67.99 & 67.05 & 59.26\\
256 & 82.38 & 79.49 & 77.13 & 68.33 & 67.52 & 57.02\\
512 & 83.43 & 81.39 & 79.12 & 69.53 & 68.35 & 58.37\\
full & 85.75 & 84.19 & 82.59 & 69.89 & 68.51 & 58.39\\
\bottomrule
\end{tabular*}
\caption{Effect of retrieval representation on classification accuracy (\%) across candidate-pool sizes. Feature-space and text-embedding retrieval are compared for Distance-KNN, Gemma3-12B, and Qwen3-8B with $k=8$. Entries report mean accuracy across 19 datasets and three seeds.}
\label{tab:embedding}
\end{table*}

\begin{figure}[!htbp]
\centering
\includegraphics[width=0.8\linewidth]{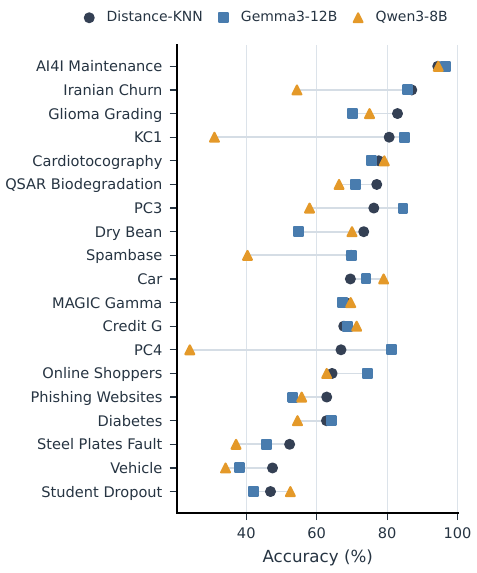}
\caption{Dataset-level classification accuracy (\%) under text-embedding retrieval at full training pools ($k=8$). Results compare Distance-KNN, Gemma3-12B, and Qwen3-8B on 19 datasets, with each value averaged over three seeds.}
\label{fig:embdatasets}
\end{figure}

At full pools, feature-based retrieval improves Distance-KNN and Qwen3-8B on all 19 datasets and Gemma3-12B on 18; embedding retrieval improves Gemma3-12B on KC1 by 2.21 percentage points. Figure~\ref{fig:embdatasets} reports the dataset-level embedding results.

\begin{table*}[t]
\centering
\captionsetup{skip=3pt}
\small
\renewcommand{\arraystretch}{1.00}
\setlength{\tabcolsep}{9pt}

\begin{tabular}{@{}lrrrrrr@{}}
\toprule
& \multicolumn{3}{c}{Gemma3-12B}
& \multicolumn{3}{c}{Qwen3-8B}\\
\cmidrule(lr){2-4}\cmidrule(lr){5-7}
Dataset & Nearest & Balanced & $\Delta \mathrm{Acc}$
        & Nearest & Balanced & $\Delta \mathrm{Acc}$\\
\midrule
AI4I Predictive Maintenance & 97.57 & 95.60 & -1.97 & 96.63 & 90.17 & -6.47\\
Car & 89.40 & 86.51 & -2.89 & 90.75 & 84.39 & -6.36\\
Cardiotocography & 90.92 & 89.51 & -1.41 & 90.45 & 84.98 & -5.48\\
Credit G & 75.00 & 69.33 & -5.67 & 74.67 & 72.67 & -2.00\\
Diabetes & 64.50 & 64.07 & -0.43 & 61.90 & 50.65 & -11.26\\
Dry Bean & 91.61 & 87.00 & -4.60 & 92.32 & 91.34 & -0.98\\
Glioma Grading & 87.30 & 83.73 & -3.57 & 86.11 & 80.95 & -5.16\\
Iranian Churn & 94.29 & 90.90 & -3.39 & 92.80 & 84.02 & -8.78\\
KC1 & 82.62 & 72.20 & -10.43 & 77.09 & 69.35 & -7.74\\
MAGIC Gamma & 84.45 & 81.42 & -3.03 & 84.98 & 80.20 & -4.78\\
Online Shoppers & 82.54 & 65.29 & -17.25 & 78.08 & 59.23 & -18.84\\
PC3 & 86.41 & 65.61 & -20.81 & 85.99 & 66.88 & -19.11\\
PC4 & 84.70 & 68.04 & -16.67 & 84.25 & 66.44 & -17.81\\
Phishing Websites & 96.93 & 95.00 & -1.93 & 87.25 & 73.96 & -13.29\\
QSAR Biodegradation & 87.74 & 85.53 & -2.20 & 88.68 & 82.70 & -5.97\\
Spambase & 90.17 & 76.79 & -13.38 & 87.06 & 71.73 & -15.33\\
Steel Plates Fault & 72.65 & 65.47 & -7.18 & 72.82 & 64.10 & -8.72\\
Student Dropout & 70.58 & 53.27 & -17.31 & 68.77 & 49.36 & -19.41\\
Vehicle & 70.20 & 61.96 & -8.24 & 68.63 & 45.10 & -23.53\\
\midrule
\rowcolor{TableBand}
Dataset mean & 84.19 & 76.70 & -7.49 & 82.59 & 72.01 & -10.58\\
\bottomrule
\end{tabular}

\caption{Effect of demonstration selection on full-pool classification accuracy (\%) for Gemma3-12B and Qwen3-8B across 19 datasets. Nearest uses the eight closest examples, while Balanced selects eight label-balanced examples from the nearest 32 candidates. Dataset entries average three seeds. $\Delta\mathrm{Acc}=\mathrm{Acc}_{\mathrm{Balanced}}-\mathrm{Acc}_{\mathrm{Nearest}}$ negative values favor nearest-neighbor selection.}
\label{tab:balance}
\end{table*}

\subsection{Label-balanced selection}
\label{app:balance}
At full pools, the nearest eight rows are compared with eight rows selected from the nearest 32 for label balance. Each step selects an available label with the fewest chosen examples; ties favor the closest remaining row. The final eight rows are ordered by distance, making label counts as even as candidate availability permits. The comparison uses seeds 42, 3407, and 5201, with matched splits, prompts, and inference settings between selection conditions within each seed.

Table~\ref{tab:balance} gives the dataset-level results behind Table~\ref{tab:balance_summary}. Balancing reduces mean accuracy by 7.49 percentage points for Gemma3-12B and 10.58 for Qwen3-8B, with decreases on every dataset. The selection changes both label composition and the identity and proximity of the retrieved rows.

\section{Complete Accuracy Results}
\label{app:additionalmetrics}
Tables~\ref{tab:allpools}--\ref{tab:allpools3} report five-seed mean accuracy (\%) for every dataset and candidate-pool size. Within each dataset and pool, the highest displayed accuracy is bold and the second-highest distinct accuracy is underlined; ties share the same emphasis. Distance-KNN denotes distance-weighted voting.

\begingroup
\makeatletter
\setlength{\@fptop}{0pt}
\setlength{\@fpbot}{0pt plus 1fil}
\makeatother
\newcommand{\APBest}[1]{\textbf{#1}}
\newcommand{\APSecond}[1]{\underline{#1}}
\newcommand{\APSetup}{%
  \normalsize
  \setlength{\tabcolsep}{2.5pt}%
  \renewcommand{\arraystretch}{0.7}%
  \setlength{\aboverulesep}{0.25ex}%
  \setlength{\belowrulesep}{0.25ex}%
  \setlength{\cmidrulesep}{0.12ex}%
}
\newcommand{\APDataset}[1]{%
  \addlinespace[3pt]%
  \multicolumn{6}{@{}l@{}}{\textbf{#1}}\\[1pt]%
}
\newcommand{\APHead}{%
  \toprule
  \textbf{Pool} &
  \textbf{Distance-KNN} &
  \textbf{Gemma3-12B} &
  \textbf{Gemma3-27B} &
  \textbf{Qwen3-8B} &
  \textbf{Qwen3-32B}\\
  \midrule
}

\clearpage
\begin{table*}[p]
\centering
\caption{Classification accuracy (\%) of Distance-KNN and four LLMs by dataset and candidate-pool size (part 1 of 3, $k=8$). Values average five seeds. Within each dataset and pool, bold and underlined values mark the highest and second-highest distinct displayed accuracies, respectively, including ties.}
\label{tab:allpools}
\APSetup
\begin{tabular*}{\textwidth}{@{\extracolsep{\fill}}lrrrrr@{}}
\APHead

\APDataset{AI4I Predictive Maintenance}
16 & \APBest{96.60} & \APSecond{94.06} & 93.72 & \APSecond{94.06} & 90.30\\
32 & \APBest{96.60} & 94.90 & 95.46 & \APSecond{95.52} & 93.42\\
64 & \APBest{96.60} & \APSecond{95.66} & 95.52 & 95.60 & 94.24\\
128 & \APBest{96.58} & 95.62 & \APSecond{95.72} & 95.46 & 94.72\\
256 & \APBest{96.54} & \APSecond{96.34} & 96.08 & 95.76 & 95.02\\
512 & \APBest{96.72} & \APSecond{96.44} & 96.38 & 95.74 & 95.86\\
full & \APBest{97.58} & \APSecond{97.56} & 97.32 & 96.78 & 97.04\\

\APDataset{Car}
16 & \APSecond{70.17} & 68.32 & \APBest{70.75} & \APBest{70.75} & 69.36\\
32 & 71.56 & 71.79 & \APBest{73.64} & \APSecond{72.37} & 72.02\\
64 & 74.68 & 74.91 & \APBest{76.65} & 74.57 & \APSecond{75.14}\\
128 & 78.84 & \APSecond{78.96} & \APBest{80.12} & 76.88 & 78.50\\
256 & \APSecond{81.39} & 81.27 & \APBest{82.77} & 80.12 & 80.46\\
512 & 84.62 & 85.32 & \APBest{86.94} & 84.16 & \APSecond{85.66}\\
full & 90.17 & 89.25 & \APBest{93.06} & 89.83 & \APSecond{90.87}\\

\APDataset{Cardiotocography}
16 & 78.03 & \APBest{81.78} & 79.81 & 80.94 & \APSecond{81.60}\\
32 & 80.94 & \APBest{84.32} & 81.97 & 82.82 & \APSecond{83.00}\\
64 & 83.29 & \APBest{84.69} & 83.94 & 82.72 & \APSecond{84.32}\\
128 & 86.85 & \APSecond{86.95} & 86.76 & 85.63 & \APBest{87.04}\\
256 & \APBest{88.45} & 87.70 & 87.89 & 87.61 & \APSecond{88.08}\\
512 & 89.39 & 89.39 & \APBest{90.42} & 88.54 & \APSecond{89.48}\\
full & 91.92 & 91.27 & \APSecond{92.02} & 90.42 & \APBest{93.52}\\

\APDataset{Credit G}
16 & 66.20 & 68.60 & 65.80 & \APSecond{69.80} & \APBest{70.00}\\
32 & \APSecond{70.80} & 68.40 & 67.20 & \APBest{72.00} & 69.40\\
64 & \APBest{70.20} & 69.00 & 67.40 & \APSecond{69.40} & 67.80\\
128 & \APSecond{70.80} & \APBest{71.40} & 69.80 & \APSecond{70.80} & 68.80\\
256 & \APBest{74.00} & 71.60 & 72.00 & \APSecond{73.00} & 68.20\\
512 & \APBest{72.20} & 70.40 & 69.80 & \APSecond{71.20} & 70.00\\
full & \APBest{73.60} & \APBest{73.60} & 73.00 & \APSecond{73.20} & 71.40\\

\APDataset{Diabetes}
16 & \APSecond{70.65} & 65.97 & 67.01 & 48.57 & \APBest{73.25}\\
32 & \APBest{71.43} & 67.01 & \APSecond{68.31} & 57.66 & \APBest{71.43}\\
64 & \APBest{70.91} & \APSecond{65.71} & 64.16 & 57.14 & \APBest{70.91}\\
128 & \APBest{75.58} & 67.53 & 70.65 & 63.12 & \APSecond{72.21}\\
256 & \APBest{74.81} & 65.71 & 69.87 & 59.22 & \APSecond{70.39}\\
512 & \APBest{74.29} & 66.23 & 70.65 & 60.78 & \APSecond{71.17}\\
full & \APBest{74.29} & 63.90 & 68.31 & 63.12 & \APSecond{69.35}\\

\APDataset{Dry Bean}
16 & 64.49 & 71.89 & \APSecond{74.39} & 70.69 & \APBest{76.04}\\
32 & 77.33 & 80.09 & \APSecond{81.60} & 79.62 & \APBest{83.08}\\
64 & 84.43 & 84.93 & \APSecond{86.20} & 85.65 & \APBest{86.27}\\
128 & 87.65 & 88.09 & \APSecond{88.21} & 87.91 & \APBest{88.22}\\
256 & \APSecond{89.24} & 89.06 & \APBest{89.71} & 89.22 & 89.09\\
512 & 90.12 & 89.52 & 89.99 & \APBest{90.31} & \APSecond{90.18}\\
full & \APSecond{92.29} & 91.78 & 92.13 & 92.20 & \APBest{92.35}\\

\APDataset{Glioma Grading}
16 & 76.43 & 75.48 & \APSecond{76.90} & 76.19 & \APBest{80.00}\\
32 & 78.10 & 78.10 & \APBest{81.43} & 75.71 & \APSecond{80.71}\\
64 & \APSecond{80.71} & 79.29 & \APBest{80.95} & 78.81 & 79.52\\
128 & 83.57 & 82.14 & \APSecond{83.81} & \APBest{84.52} & \APSecond{83.81}\\
256 & \APSecond{83.10} & 81.43 & 80.95 & 80.71 & \APBest{83.57}\\
512 & \APSecond{84.29} & 83.81 & \APBest{84.52} & 83.57 & 82.62\\
full & 84.52 & \APSecond{85.24} & \APBest{86.19} & 84.29 & 83.33\\

\bottomrule
\end{tabular*}
\end{table*}

\clearpage
\begin{table*}[t]
\centering
\caption{Classification accuracy (\%) of Distance-KNN and four LLMs by dataset and candidate-pool size (part 2 of 3, $k=8$). Values average five seeds. Within each dataset and pool, bold and underlined values mark the highest and second-highest distinct displayed accuracies, respectively, including ties.}
\label{tab:allpools2}
\APSetup
\begin{tabular*}{\textwidth}{@{\extracolsep{\fill}}lrrrrr@{}}
\APHead

\APDataset{Iranian Churn}
16 & \APSecond{84.13} & 82.29 & \APBest{84.32} & 73.65 & 81.84\\
32 & 84.70 & \APSecond{86.35} & \APBest{87.94} & 78.98 & 83.75\\
64 & 85.14 & \APSecond{86.29} & \APBest{87.81} & 81.78 & 85.65\\
128 & 86.98 & 87.56 & \APSecond{88.00} & 83.81 & \APBest{88.57}\\
256 & 89.14 & 87.62 & \APSecond{89.27} & 82.92 & \APBest{90.60}\\
512 & \APSecond{92.00} & 90.92 & \APBest{92.51} & 87.56 & 91.81\\
full & \APSecond{94.35} & 93.71 & \APBest{94.79} & 91.75 & 94.10\\

\APDataset{KC1}
16 & \APBest{84.64} & 73.93 & \APSecond{81.90} & 71.66 & 77.35\\
32 & \APBest{83.41} & 73.74 & \APSecond{81.52} & 64.27 & 75.92\\
64 & \APBest{82.37} & 73.27 & \APSecond{81.99} & 66.45 & 74.60\\
128 & \APBest{83.79} & 73.46 & \APSecond{81.42} & 68.44 & 74.79\\
256 & \APBest{83.89} & 75.26 & \APSecond{80.38} & 67.49 & 76.68\\
512 & \APBest{83.60} & 77.73 & \APSecond{82.94} & 70.62 & 77.25\\
full & \APBest{86.82} & 80.57 & \APSecond{85.59} & 75.26 & 80.38\\

\APDataset{MAGIC Gamma}
16 & 68.69 & 65.95 & 67.14 & \APSecond{69.01} & \APBest{69.87}\\
32 & 69.91 & 73.14 & \APBest{73.21} & 71.48 & \APSecond{73.15}\\
64 & 74.41 & \APBest{76.66} & \APSecond{76.60} & 75.15 & 75.81\\
128 & 76.56 & \APBest{77.88} & \APSecond{77.15} & 77.06 & 76.69\\
256 & 78.35 & 78.23 & \APSecond{78.39} & \APBest{78.44} & 77.55\\
512 & 79.32 & 79.12 & \APSecond{79.40} & \APBest{79.54} & 78.58\\
full & 84.54 & 84.22 & \APBest{84.71} & \APSecond{84.64} & 82.64\\

\APDataset{Online Shoppers}
16 & \APBest{84.51} & 75.26 & \APSecond{81.75} & 73.53 & 81.46\\
32 & \APBest{85.81} & 76.03 & \APSecond{82.79} & 74.06 & 80.54\\
64 & \APBest{86.12} & 75.13 & \APSecond{82.68} & 72.02 & 78.64\\
128 & \APBest{87.01} & 77.92 & \APSecond{83.83} & 74.53 & 79.16\\
256 & \APBest{87.09} & 78.90 & \APSecond{84.66} & 74.97 & 78.80\\
512 & \APBest{87.32} & 79.11 & \APSecond{85.04} & 76.35 & 79.53\\
full & \APBest{87.98} & 82.32 & \APSecond{86.08} & 78.02 & 78.98\\

\APDataset{PC3}
16 & \APBest{89.94} & 72.10 & \APSecond{79.75} & 72.48 & 72.61\\
32 & \APBest{89.94} & 77.96 & \APSecond{81.27} & 79.36 & 77.71\\
64 & \APBest{89.55} & 76.82 & \APSecond{81.15} & 78.34 & 77.83\\
128 & \APBest{89.43} & 77.71 & \APSecond{82.04} & 78.98 & 79.24\\
256 & \APBest{89.68} & 81.78 & \APSecond{85.73} & 81.78 & 82.42\\
512 & \APBest{88.79} & 82.42 & \APSecond{86.50} & 82.17 & 84.08\\
full & \APBest{89.30} & 85.35 & \APSecond{88.54} & 85.35 & 86.24\\

\APDataset{PC4}
16 & \APBest{87.67} & 69.32 & \APSecond{76.30} & 67.12 & 72.60\\
32 & \APBest{87.53} & 70.82 & \APSecond{78.36} & 72.60 & 74.38\\
64 & \APBest{87.81} & 73.97 & \APSecond{77.53} & 71.37 & 75.62\\
128 & \APBest{87.53} & 75.62 & \APSecond{81.23} & 74.11 & 75.75\\
256 & \APBest{88.22} & 78.08 & \APSecond{81.78} & 75.48 & 79.18\\
512 & \APBest{88.08} & 78.77 & \APSecond{83.84} & 77.67 & 82.60\\
full & \APBest{88.49} & 84.11 & \APSecond{87.40} & 82.88 & 84.79\\

\bottomrule
\end{tabular*}
\end{table*}

\clearpage
\begin{table*}[t]
\centering
\caption{Classification accuracy (\%) of Distance-KNN and four LLMs by dataset and candidate-pool size (part 3 of 3, $k=8$). Values average five seeds. Within each dataset and pool, bold and underlined values mark the highest and second-highest distinct displayed accuracies, respectively, including ties.}
\label{tab:allpools3}
\APSetup
\begin{tabular*}{\textwidth}{@{\extracolsep{\fill}}lrrrrr@{}}
\APHead

\APDataset{Phishing Websites}
16 & 74.32 & \APBest{79.64} & 77.79 & 56.98 & \APSecond{79.62}\\
32 & 82.95 & \APBest{84.09} & 83.74 & 59.11 & \APSecond{83.78}\\
64 & \APSecond{85.75} & 85.66 & \APBest{86.06} & 62.15 & 85.55\\
128 & \APBest{89.19} & 88.12 & 88.73 & 65.17 & \APSecond{88.79}\\
256 & \APBest{90.25} & 89.24 & \APSecond{90.16} & 69.46 & 89.58\\
512 & \APSecond{91.63} & 91.03 & \APBest{91.92} & 73.04 & 91.32\\
full & \APSecond{96.93} & \APBest{97.00} & 96.73 & 86.71 & \APSecond{96.93}\\

\APDataset{QSAR Biodegradation}
16 & \APBest{70.57} & 66.60 & 65.09 & \APSecond{69.06} & 67.17\\
32 & \APBest{73.21} & 69.81 & 70.38 & \APSecond{72.83} & 70.75\\
64 & 77.17 & 76.60 & \APBest{78.87} & 76.23 & \APSecond{78.49}\\
128 & \APSecond{80.19} & 79.06 & 80.00 & \APSecond{80.19} & \APBest{80.38}\\
256 & \APSecond{83.02} & 81.32 & 81.89 & \APBest{83.40} & 82.26\\
512 & \APBest{83.58} & \APSecond{83.21} & \APBest{83.58} & \APSecond{83.21} & 82.64\\
full & \APSecond{85.28} & 84.34 & \APSecond{85.28} & \APBest{85.66} & 84.72\\

\APDataset{Spambase}
16 & 73.19 & 68.89 & \APSecond{74.19} & 58.05 & \APBest{78.79}\\
32 & \APSecond{75.92} & 71.80 & 73.28 & 66.98 & \APBest{78.87}\\
64 & \APSecond{80.61} & 74.92 & 76.70 & 68.03 & \APBest{81.39}\\
128 & 81.52 & 81.30 & \APSecond{82.73} & 78.09 & \APBest{84.60}\\
256 & \APSecond{85.47} & 83.99 & 84.12 & 81.56 & \APBest{86.25}\\
512 & \APSecond{86.42} & 84.21 & 85.03 & 82.30 & \APBest{86.98}\\
full & \APSecond{91.28} & 90.54 & 90.11 & 87.20 & \APBest{91.63}\\

\APDataset{Steel Plates Fault}
16 & 44.62 & \APBest{49.54} & 46.97 & 48.21 & \APSecond{49.44}\\
32 & 49.85 & \APSecond{51.28} & \APBest{52.72} & 49.33 & 51.18\\
64 & 52.82 & 52.72 & \APSecond{54.05} & 52.92 & \APBest{54.46}\\
128 & 57.74 & 58.77 & \APSecond{59.28} & 56.10 & \APBest{59.90}\\
256 & 61.74 & 62.87 & \APSecond{63.49} & 61.64 & \APBest{64.62}\\
512 & 67.49 & \APBest{69.23} & 67.08 & 66.05 & \APSecond{68.72}\\
full & 75.18 & 73.85 & \APSecond{75.38} & 74.46 & \APBest{75.59}\\

\APDataset{Student Dropout}
16 & 54.99 & \APSecond{62.75} & \APBest{69.39} & 53.14 & 46.64\\
32 & 60.41 & \APSecond{62.26} & \APBest{68.85} & 54.27 & 48.31\\
64 & 63.25 & \APSecond{63.84} & \APBest{69.39} & 58.24 & 50.38\\
128 & \APSecond{65.19} & 64.51 & \APBest{70.20} & 61.40 & 54.45\\
256 & \APSecond{65.33} & 64.97 & \APBest{69.80} & 62.03 & 56.12\\
512 & \APSecond{66.73} & 66.14 & \APBest{70.47} & 64.70 & 57.79\\
full & \APSecond{70.74} & 70.56 & \APBest{73.18} & 68.89 & 64.15\\

\APDataset{Vehicle}
16 & 43.53 & \APSecond{46.82} & \APBest{48.94} & 38.12 & 43.53\\
32 & 48.94 & \APBest{54.12} & 52.00 & 39.76 & \APSecond{52.24}\\
64 & 55.29 & \APBest{59.29} & \APSecond{58.59} & 49.65 & 58.35\\
128 & \APSecond{63.06} & \APBest{64.00} & 62.82 & 58.35 & 62.82\\
256 & 66.59 & 66.12 & \APSecond{67.53} & 63.29 & \APBest{68.24}\\
512 & \APBest{70.35} & \APSecond{69.41} & \APSecond{69.41} & 68.47 & 68.94\\
full & 69.18 & 68.71 & \APBest{70.12} & 66.82 & \APSecond{69.41}\\

\bottomrule
\end{tabular*}
\end{table*}
\clearpage
\endgroup

\end{document}